\documentstyle[12pt,epsf]{article}
\begin{document}

\title{Conductivity activation energy for bilayer heterostructures at
integer filling factors}
\author{V.Falko, S.V.Iordanski and A.Kashuba \\ {\it Lancaster University
School of Physics, UK} \\ {\it Landau Institute for Theoretical Physics,
Moscow 142432}}



\maketitle

Integer filling factors of a hetero-structure 2D electron gas (2DEG) in an
external magnetic field are special ones because a huge degeneracy of the
ground state is gone here. It justifies the Hartree-Fock approximation with
the accuracy limited only by normally a small parameter: $V^{int}/\hbar
\omega_0$, where $V^{int}$ is the energy of the Coulomb interaction and $
\omega_0$ is the frequency of the cyclotron resonance. Such an approach predicts
the ground state of a single layer 2DEG at $\nu=1$ to be a ferromagnet with
the degenerate total spin orientation. The elementary excitations of 2DEG are
electron-hole pairs or excitons and in the close binding limit of vanishing
momentum they transform into the elementary excitations of a ferromagnet -
spin-waves. The latter are gapless \cite{BIE82} and do not interact with each
other \cite{LL82} if Zeeman energy is neglected - the two consequences of the
Goldstone theorem. In the limit of large momentum the electron and the hole of
an exciton are well separated and they become the elementary charged
excitations. In the Ref.\cite{SKKR93} it was shown that a special topological
spin texture in a 2D ferromagnet called a skyrmion \cite{BP75} carries a unit
charge while costing only half the exciton's energy to appear.

The case of a bilayer 2DEG turned out to be a more rich one. The Hartree-Fock
approximation does not apply here except for two limiting cases. The first one
is the case of well separated layers which is a common setup in the experiment
\cite{Exper,K00} and where, theoretically, one starts from the two single
layer ferromagnets and makes the perturbation expansion in powers of
interlayer interactions \cite{SSZ97}. And the second one is the symmetric case
defined in such a way that one can freely rotates an electron spinor in both
layer and spin spaces. The latter requires to approximate the Coulomb
interaction by its symmetric part and to neglect all symmetry breaking fields
like Zeeman energy. The first attempts in this direction dealt with the case
of filling factor $\nu=1$ and relied heavily on the assumption of a saturated
spin polarization of electrons \cite{RHV86,BI87}. This symmetric approximation
turned out to be useful to determine the exciton energy in bilayer
\cite{BI87}. Recent works Refs.\cite{PhaseDiagram,DD99} specialize to the
bilayer heterostructure case $\nu=2$, employ the Hartree- Fock approximation
and predict a phase diagram that features three phases: the ferromagnetic, the
canted antiferromagnetic and a special spin-singlet phase. In this paper we
reproduce the phase diagram of the Refs.\cite{PhaseDiagram, DD99} isolating
the symmetric and the symmetry breaking parts of the Hamiltonian in a
consistent way. Our approach reveals the Hartree-Fock phase diagram to be
indeed exact in the limit $V^{anis}/V^{sym}\rightarrow 0$, where $V^{sym}$ is
the $SU(4)$-symmetric part of the bilayer Hamiltonian whereas $V^{anis}$ is
anisotropy interactions that reduce the bilayer Hamiltonian symmetry to
$SU(2)\otimes SU(2)$. We prove the stability of all phases with respect to
long-range spatial perturbations.

Our new results concern the energy of topological excitations in bilayers. We
find that low-energy excitations over the bilayer ground state is governed by
the $U(4)/U(\nu)\otimes U(4-\nu)$ coset non-linear Sigma Model. We identify
charged excitations in the bilayer with skyrmions or topological excitations
of this non-linear Sigma Model \cite{BP75,SKKR93,IPF99}. We calculate the
skyrmion energy gap to vary dramatically over the bilayer parameter space and
we find a sharp dip of this gap in the canted antiferromagnetic phase of the
bilayer. Our work was motivated by recent measurements of the diagonal
conductivity activation energy \cite{K00}. In this paper we suggest to
identify the experimental activation energy with the energy gap of skyrmion,
and we have got a qualitative agreements with the results of Ref.\cite{K00}
along this line.

\section*{Hamiltonian of 2DEG bilayer}

The electronic Hamiltonain of a 2DEG in a confining potential $V(\vec{\rho})$
and in an external magnetic field $H$ consists of a one-particle part as well
as a Coulomb interaction part:
\begin{equation}\label{basicHam}  \begin{array}{c}\displaystyle H=
\int\psi^{+}_\alpha(\vec{\rho})\left({1\over 2m}\left[-i\vec{\nabla}+\vec{A}
(\vec{\rho})\right]^2+V(\vec{\rho})-|g|\mu_BH\sigma^z_{\alpha\beta}\right)
\psi_\beta(\vec{\rho})\,d^3\vec{\rho}\ + \nonumber\\ \displaystyle {1\over
2}\int\int{e^2\over |\vec{\rho}-\vec{\rho'}|}\psi^{+}_\alpha(\vec{\rho})
\psi^{+}_\beta(\vec{\rho'})\psi_\beta(\vec{\rho'})\psi_\alpha(\vec{\rho})\,
d^3\vec{\rho}\ d^3\vec{\rho'}, \end{array} \end{equation} where $\alpha,\beta=
\pm$ are spin indices and thereafter a sum over repeated indices is implied.
We use such units that $\hbar=1$, $e=c$ and $H=B=1$. The latter implies that
all distances can be expressed in terms of the so-called magnetic length:
$l_H=\sqrt{c\hbar/eH}=1$. We split three coordinates $\vec{\rho}$ into a
perpendicular to the layer coordinate $\xi$ and two in-plane coordinates
$\vec{r}=(x,y)=(z,\bar{z})$. We assume that the confining potential is uniform
over the plane: $V(\vec{\rho})=V(\xi)$, and represents a double well structure
in the transverse direction as shown on the Fig.1, with the two wells being
separated by the distance $d$. We use only two eigen functions: the lowest
energy symmetric $\chi_S(\xi)$ and antisymmetric $\chi_A(\xi)$, from a set of
one-electron eigen functions in the confining potential $V(\xi)$ and we expand
an electron second-quantized operator in terms of these two eigen functions:
\begin{figure}\label{Bilayer}
\epsfxsize=4.0in             \centerline{\epsffile{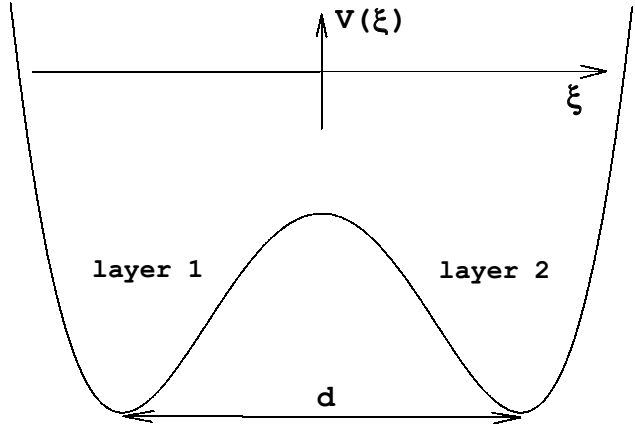}}
\caption{Schematic view of a confining potential $V(\xi)$ in a typical
Bilayer setup}                   \end{figure}
\begin{equation}\label{EigenFunc}
\psi_\alpha(\vec{\rho})=\sum_{\tau,n,p}\chi_\tau(\xi)\phi_{n,p}(\vec{r})
c_{n\alpha\tau p}, \end{equation} where $c^+_{\alpha\tau p}$ and
$c_{\alpha\tau p}$ are electron creation and anhilation operators,
$\phi_{n,p}(z\bar{z})$ is an electron wave function number $p$ in the Landau
gauge in the $n$'s Landau Level, the index $\tau=1,2$ being the layer index
and the layer wave functions read:
\begin{equation}\label{SymAsym}
\chi_{1,2}(\xi)={\chi_S(\xi)\pm\chi_A(\xi)\over \sqrt{2}}.
\end{equation} We restrict our model to the case of a sufficiently strong
magnetic field, such that the cyclotron energy $\hbar\omega_0$ dominates over
the Coulomb, Zeeman and the level splitting: $E_A-E_S$, energies. Thus, we
specialize to the lowest Landau Level and retain only the term $n=0$ in
(\ref{EigenFunc}).

Plugging the wave function (\ref{EigenFunc}) into the Hamiltonian
(\ref{basicHam}) we find a 2DEG Hamiltonian as:
\begin{equation}\label{reducedHam}
\begin{array}{c} \displaystyle
H\ =\ {1\over 2m}c^+_{\alpha\tau p}c_{ \alpha\tau p}-
c^+_{\alpha\tau_1p}\left(t\tau^x_{\tau_1\tau_2}+\mu^z\tau^z_{\tau_1\tau_2}
\right)c_{\alpha\tau_2p}-|g|\mu_BH\,c^+_{\alpha\tau p}\sigma^z_{\alpha\beta}
c_{\beta\tau p}+ \\ \displaystyle +\ {1\over 2}\sum_{p_1..p_4}\int\int
d^2\vec{r}d^2\vec{r'}\ V^{\tau_1\tau_4}_{\tau_2\tau_3}(\vec{r}-\vec{r'})
\phi^*_{p_1}(\vec{r})\phi^*_{p_2}(\vec{r'})\phi_{p_3}(\vec{r'})\phi_{p_4}
(\vec{r})\\ \displaystyle c^+_{\alpha\tau_1p_1}c^+_{\beta\tau_2p_2}
c_{\beta\tau_3p_3} c_{\alpha\tau_4p_4}, \end{array} \end{equation} where we
have defined a hopping constant:
\begin{equation}\label{Hopt} t\ ={1\over 2}\int\int d^2\vec{r}d\xi\
\phi^*_{p}(\vec{r})\chi_{\tau_1}(\xi)\tau^x_{\tau_1\tau_2}V(\xi)\chi_{\tau_2}
(\xi)\phi_{p}(\vec{r}), \end{equation} as well as an external electrostatic
potential created by an asymmetric gate charge:
\begin{equation}\label{GateMuz}\mu^z\ ={1\over 2}\int\int d^2\vec{r}d\xi\
\phi^*_{p}(\vec{r})\chi_{\tau_1}(\xi)\tau^z_{\tau_1\tau_2}V(\xi)\chi_{\tau_2}
(\xi)\phi_{p}(\vec{r}), \end{equation} whereas the Coulomb interaction matrix
reads:
\begin{equation}\label{CoulOper}V^{\tau_1\tau_4}_{\tau_2\tau_3}(\vec{r}-
\vec{r'})=\int\int{\chi_{\tau_1}(\xi)\chi_{\tau_2}(\xi')\chi_{\tau_3}(\xi')
\chi_{\tau_4}(\xi)\over\sqrt{(\xi-\xi')^2+\left(\vec{r}-\vec{r'}\right)^2}}\
d\xi d\xi'. \end{equation} We use notations: $\tau^x$, $\tau^y$ and $\tau^z$,
for the Pauli matrices in the layer space whereas we use notations:
$\sigma^x$, $\sigma^y$ and $\sigma^z$, for the Pauli matrices in the spin
space. The hopping constant can be related to the splitting of the symmetric
and the antisymmetric levels: $t=E_A-E_S$. The electrostatic potential
$\mu^z$, which can be viewed as a difference between the chemical potentials
in the two layers, breaks down the symmetry between the two wells of $V(\xi)$
potential. This term appears naturally when a single gate is fabricated to
control the electron density in the bilayer. In the limit $d\rightarrow 0$,
$\mu^z$ vanishes too, whereas in the limit of large layer separation: $d
\rightarrow\infty$, $\mu^z\rightarrow\infty$ and electrons reside only on the
layer adjacent to the gate. We assume that the energy of a capacitor formed by
the two layers is much lower than the characteristic Coulomb energy $e^2/
\kappa l_H^3$, per area, where $\kappa$ is the dielectric constant. We note
the invariance of the Coulomb energy: (\ref{CoulOper}) under the following
transformations: $\tau_1\leftrightarrow\tau_4$, $\tau_2\leftrightarrow\tau_3$
as well as $(\tau_1\tau_4)\leftrightarrow(\tau_2\tau_3)$. To fully exploit
these symmetries we cast the Eq.\ref{CoulOper} into a more suitable
representation:
\begin{equation}\label{CoulSymm}
V^{\tau_1\tau_4}_{\tau_2\tau_3}(\vec{r}-\vec{r'})=V^{\mu\nu}(\vec{r}-\vec{r'})
\tau^\mu_{\tau_1\tau_4}\tau^\nu_{\tau_2\tau_3},      \end{equation}
where $\tau^0$ is the unit matrix, $V^{\mu\nu}$ is a $3\times 3$ symmetric
interaction matrix with indices $\mu$, $\nu$ running over a set $(0,z,x)$. If
there is a symmetry of the Coulomb interaction under an exchange of layers:
$(\xi\xi')\leftrightarrow (-\xi-\xi')$ and $1\leftrightarrow 2$ then it
restricts further values of the interaction matrix: $V^{0z}=0$ and $V^{zx}=0$.
But in the presence of a gate asymmetry we shall keep the matrix element:
$V^{0z}$. Therefore, the Coulomb interaction matrix for symmetric bilayer
2DEG depends on four parameters: $V^{00}>0$, $V^{0x}$, $V^{xx}>0$,
$V^{zz}>0$. We note also, that $V^{0x}\sim\chi$, $V^{xx}\sim\chi^2$, whereas
$V^{zz}\sim d^2/|z|^3$ as $|z|\rightarrow\infty$. In the following we shall
neglect $V^{xx}$ matrix element.

Next, we split the total bilayer Hamiltonian (\ref{reducedHam}) into two
parts: the first one contains a dominant Coulomb energy term:
\begin{equation}\label{symHam}
\begin{array}{c} \displaystyle
H^{sym}\ =\ {1\over 2m}c^+_{\alpha\tau p}c_{\alpha\tau p}+{1\over
2}\int{d^2\vec{q}\over (2\pi)^2} V^{00}(\vec{q}) N(\vec{q})N(-\vec{q}),
\end{array} \end{equation} where $V^{\mu\nu}(\vec{q})$ is the Fourier
transform of $V^{\mu\nu}$ $(\vec{r})$ multiplied by a factor 
$\exp(-\vec{q}^2/4)$ and the electron density operator reads:
\begin{equation}\label{N0}
N(\vec{q})\ =\ \sum_p\ c^+_{\alpha\tau p}c_{\alpha\tau p-q_y}
\exp{-iq_x(p-{q_y\over 2})} \end{equation} This part of the Hamiltonian is
invariant under uniform rotations from the $SU(4)$ Lee group in the combined
spin and layer space. Every of its eigen energy is hugely degenerate. Given
any eigen state $|\Psi\rangle_0$ a set of related eigen states can be
generated by applying rotations: $|\Psi\rangle=U|\Psi\rangle_0$, where $U\in
SU(4)$. For Landau level filling factor $\nu=1$, $\nu=2$ and $\nu=3$ we assume
that the the bilayer ground state is uniform over $p$-orbitals:
\begin{equation}\label{PsiRitor}
\Psi\ =\ \prod_{i=1}^{\nu}\prod_pc^+_{\alpha_i\tau_i p}\ |empty\rangle ,
\end{equation}
and we prove in the next section that this state is stable with respect to
long-range spatial perturbations. One can easily check by inspection that any
such wave-function (\ref{PsiRitor}) represents an eigen function of the
$H^{sym}$ (\ref{symHam}). The remaining few terms in the Hamiltonian
(\ref{reducedHam}) are treated like perturbations:
\begin{equation}\label{anisHam}
\begin{array}{c} \displaystyle
H^{anis}\ =\ -c^+_{\alpha\tau_1p}\left(t\tau^x_{\tau_1\tau_2}+\mu^z
\tau^z_{\tau_1\tau_2}\right)c_{\alpha\tau_2p}-|g|\mu_BH\,c^+_{\alpha\tau p}
\sigma^z_{\alpha\beta}c_{\beta\tau p}\ +     \nonumber\\  \displaystyle
+\ {1\over 2}\int{d^2\vec{q}\over (2\pi)^2} V^{\mu\nu}(\vec{q})
T^\mu(\vec{q})T^\nu(-\vec{q}),
\end{array}       \end{equation} where (see e.g. \cite{BI87})
\begin{equation}\label{Txyz}
T^\mu(\vec{q})\ =\ \sum_p\ c^+_{\alpha\tau_1p}\tau^\mu_{\tau_1\tau_2}
c_{\alpha\tau_2p-q_y}\exp{-iq_x(p-{q_y\over 2})} \end{equation} with
$(\mu\nu)\ne(00)$. The Hamiltonian (\ref{anisHam}) breaks down the $SU(4)$
symmetry but it is still invariant under separate rotations in the spin and
layer space: $SU(2)\otimes SU(2)$. We shall call this part of the Hamiltonian
the anisotropy Hamiltonian. It lifts the degeneracy of eigen states of the
$SU(4)$-symmetric Hamiltonian (\ref{symHam}). An important point to note here
is that a splitting of energy levels is determined by matrix elements of weak
anisotropy Hamiltonian (\ref{anisHam}) truncated to a linear space of the
symmetric Hamiltonian (\ref{symHam}) level degeneracy. There are no
Fermi-liquid type renormalizations of the constants of the anisotropy
Hamiltonian (\ref{anisHam}) due to the $SU(4)$-symmetric Hamiltonian
(\ref{symHam}). In other word the mean-field Hartree-Fock approach is perfect
for the $\nu=1$, $\nu=2$ and $\nu=3$ cases.

Our guiding analogy in treating the total bilayer Hamiltonian (\ref{symHam},
\ref{anisHam}) lies in the theory of magnetism. We will see below that there
exists a local order parameter: $Q$, very much like magnetization. And we aim
to express the total bilayer Hamiltonian (\ref{symHam},\ref{anisHam}) in terms
of this order parameter $Q$. The exchange-like Hamiltonian (\ref{symHam}) has
to be expanded in powers of spatial variations of order parameter $Q(\vec{r})$
with the second power of gradients being the important contribution, whereas
only locally homogeneous $Q$ has to be retain in the anisotropy Hamiltonian
(\ref{anisHam}). In the next section we carry out the first step whereas in
the next-to-next section we transform the anisotropy energy.

\section*{SU(4) Symmetric Case}

In this section we specialize to the $SU(4)$-symmetric part of the bilayer
2DEG Hamiltonian (\ref{symHam}) which is invariant under the global rotations
of a four component electron spinor by the $4\times 4$ matrix $U$ from the
$SU(4)$ Lee group. We find it useful the ground state of operators $c$ to be
the reference state. Any non-homogeneous state is then generated by rotation:
$U(t,\vec{r})$. And the action of bilayer is some functional of it:
\begin{equation}\label{ActLagr}
S[U(t,\vec{r})]=-i\ \mbox{tr}\log\int{\cal D}c^+(t){\cal D}c(t)\exp\left(
i\int{\cal L}dt\right) \end{equation} where the symmetric Lagrangian of
bilayer 2DEG reads:
\begin{equation}\label{GradHam} \begin{array}{c}\displaystyle
{\cal L}=\int c^+_{\alpha\tau_1}\left[i{\partial\over\partial t}\ -\ {1\over
2m}\left(-i\vec{\nabla}+\vec{A}_0+\vec{\Omega}_{\alpha\beta\tau_1\tau_2}
\right)^2\right]c_{\beta\tau_2}\ d^2\vec{r}\ +\\ \displaystyle +\ \int
c^+_{\alpha\tau_1}\Omega^t_{\alpha\beta\tau_1\tau_2}c_{\beta\tau_2}\ d^2
\vec{r}\ -\ {1\over 2}\int{d^2\vec{q}\over (2\pi)^2}V^{00}(\vec{q})N(\vec{q})
N(-\vec{q}), \end{array} \end{equation} where $\vec{A}_0=(0,A^y)$ is the
vector potential of the external magnetic field and
\begin{equation}\label{UrotS}\Omega^t=i\,U^+{\partial\over\partial t}U,
\ \ \ \vec{\Omega}=-i\,U^+\vec{\nabla}U. \end{equation}
In the limit of slow spatial variations of rotation $U(t,\vec{r})$ this
functional has an expansion in powers of $\Omega$. In this context the
functional (\ref{ActLagr}) is called an effective low-energy Goldstone Action,
and we are going to find it in this section. All calculations follows step in
step those done in Ref.\cite{IPF99} for the case of a single layer, and here
we emphasized only the points of difference.

Calculations of the Goldstone Action can be carried through for three filling
factors of the bilayer 2DEG: $\nu=1$, $\nu=2$ and $\nu=3$, at once. We start
with choosing the reference state: i) $n=1$, one electron with spin up fills
every orbital $p$ of the lowest Landau Level of the first layer:
$|+\rangle_1$; ii) $n=2$, two electrons - one with spin up on the first layer
and the other with spin down on the second layer - fill every orbital $p$ the
lowest Landau Level: $|+\rangle_1|-\rangle_2$; iii) three electrons
- one with spin down on the first layer and the two other with spin up and
down on the second layer - fill every orbital $p$ the lowest Landau Level:
$|-\rangle_1|+\rangle_2|-\rangle_2$. The case iii) reduces to the case i) if
one makes the electron-hole transformation.

The one-electron Green functions defined for the reference state of the
Hamiltonian (\ref{GradHam}) in the homogeneous limit: $\vec{\Omega}=0$, reads:
\begin{equation}\label{GreenF0} \begin{array}{c} \displaystyle
G^0_{\alpha\tau_1,\beta\tau_2}(\epsilon)={1\over2}\left(\Sigma^0_{\alpha\beta
,\tau_1\tau_2}+\Sigma^z_{\alpha\beta,\tau_1\tau_2}\right){1\over\epsilon+
E_0-\mu-i0}+ \\ \displaystyle +{1\over2}\left(\Sigma^0_{\alpha\beta,
\tau_1\tau_2}-\Sigma^z_{\alpha\beta,\tau_1\tau_2}\right){1\over\epsilon-
\mu+i0},   \end{array}  \end{equation} where $\mu$ is the chemical potential,
\begin{equation}\label{S0123} \Sigma^0_{\alpha\beta,\tau_1\tau_2}=\sigma^0_{
\alpha\beta}\tau^0_{\tau_1\tau_2}=\left( \begin{array}{cccc}
    1 & 0 & 0 & 0  \\
    0 & 1 & 0 & 0  \\
    0 & 0 & 1 & 0  \\
    0 & 0 & 0 & 1  \\
\end{array} \right), \end{equation} and in the case $n=1,3$:
\begin{equation}\label{Sz13}
\Sigma^z_{\alpha\beta,\tau_1\tau_2}= \pm\left( \begin{array}{cccc}
    1 & 0 & 0 & 0  \\
    0 & -1 & 0 & 0  \\
    0 & 0 & -1 & 0  \\
    0 & 0 & 0 & -1  \\
\end{array} \right), \end{equation} whereas in the case $n=2$:
\begin{equation}\label{Sz2}
\Sigma^z_{\alpha\beta,\tau_1\tau_2}=\left( \begin{array}{cccc}
    1 & 0 & 0 & 0  \\
    0 & 1 & 0 & 0  \\
    0 & 0 & -1 & 0  \\
    0 & 0 & 0 & -1  \\
\end{array} \right). \end{equation}
\begin{figure}\label{Diagrams}
\epsfxsize=4.0in
\centerline{\epsffile{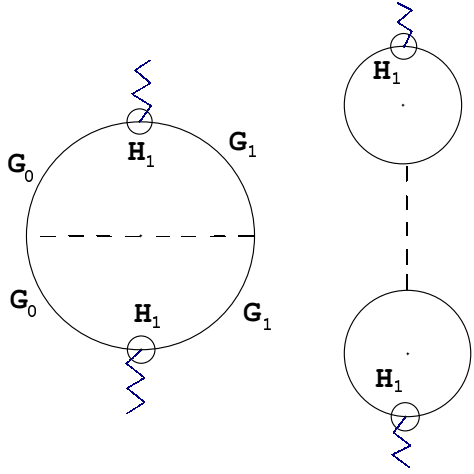}}
\caption{Second order Hartree-Fock diagrams}
\end{figure} The effective action can be written as:
\begin{equation}\label{EffectAct} S=S_0+S_2, \end{equation} where
\begin{equation}\label{Action0}S_0=i\,\mbox{tr}\log{G\over G_0}\end{equation}
and $S_2$ is conveniently represented in terms of diagrams in the Fig.2. As it
is explained in the Ref.\cite{IPF99} the first order perturbation correction
to the Green function: $\delta G=G-G_0$, contains an electron propagation on
the first excited Landau Level. Thus, we also need the bare Green function of
an excited electron on this level:
\begin{equation}\label{GreenF1} \begin{array}{c} \displaystyle
G^1_{\alpha\tau_1,\beta\tau_2}(\epsilon)={1\over2}\left(\Sigma^0_{\alpha\beta
,\tau_1\tau_2}+\Sigma^z_{\alpha\beta,\tau_1\tau_2}\right){1\over\epsilon-1/m+
E_1-\mu+i0}+ \\\displaystyle +{1\over2}\left(\Sigma^0_{\alpha\beta,
\tau_1\tau_2}-\Sigma^z_{\alpha\beta,\tau_1\tau_2}\right){1\over\epsilon-1/m-
\mu+i0}, \end{array} \end{equation} where
\begin{equation}\label{E0E1} E_0=2E_1=\sqrt{\pi\over 2}{e^2\over\kappa
l_H}. \end{equation} As it is explained in the Ref.\cite{IPF99} the gradient
vector field enters the following terms in the Hamiltonian
\begin{equation}\label{H1} H_1={1\over 2m}\int c^+_{\alpha\tau_1}\left(
\Omega^+_{\alpha\beta,\tau_1\tau_2}\hat{\Pi}_-+\Omega^-_{\alpha\beta,\tau_1
\tau_2}\hat{\Pi}_+\right) c_{\beta\tau_2}\ d^2\vec{r}, \end{equation}
\begin{equation}\label{H2} H_2={1\over 2m}\int c^+_{\alpha\tau_1}\left[
\left(\vec{\Omega}^2\right)_{\alpha\beta,\tau_1\tau_2}-i\left(\partial_\mu
\Omega_\mu\right)_{\alpha\beta,\tau_1\tau_2}\right]c_{\beta\tau_2}\
d^2\vec{r}, \end{equation} where the operators $\hat{\Pi}_\pm$ shift an electron
between the adjacent Landau Levels:
\begin{equation}\label{PiOper}\hat{\Pi}_-\phi_{np}(\vec{r})=\sqrt{2n}
\phi_{n-1p}(\vec{r}), \ \ \ \hat{\Pi}_+\phi_{n-1p}(\vec{r})=\sqrt{2n}
\phi_{np}(\vec{r}), \end{equation} though only the $n=0$ and $n=1$ Landau Level
states are relevant for our problem. The two new gradient vector fields in the
Hamiltonian (\ref{H1}) are defined as follows:
\begin{equation}\label{OmegaPM}\Omega_\pm=-i\,U^+\left(\partial_y\mp i
\partial_x\right)U, \end{equation} These two components are real:
$\Omega_-=\Omega_+^*$. An expansion of the 2DEG action up to the second power
of the Hamiltonian (\ref{H1}) reads:
\begin{equation}\label{ActExp0}
\delta S_0=i\mbox{tr}\left(H_1G_0\right)+{i\over 2}\mbox{tr}\left(H_1G_0H_1
G_0\right)+i\mbox{tr}\left(H_2G_0\right). \end{equation} Combining this
expansion with diagrams of the Fig.2 we find the low-energy Goldstone
Hamiltonian as follows:
\begin{equation}\label{GoldHamGeneral} H_G={E_1\over 8}\int\ {d^2\vec{r}
\over 2\pi}\ \mbox{tr}\left((\Sigma^0-\Sigma^z)\Omega_-(\vec{r})(\Sigma^0+
\Sigma^z)\Omega_+(\vec{r})\right)\end{equation} The insertion matrices in
(\ref{GoldHamGeneral}) are non-negative diagonal ones and they represent the
occupation number for the electron states:
\begin{equation} \label{Nfilled} N={1\over 2}\left(\Sigma^0+\Sigma^z\right)=
\left(  \begin{array}{c|c}    1 & 0 \\   \hline    0 & 0
\end{array} \right) \end{equation} and
\begin{equation} \label{Nempty} \Sigma^0-N={1\over 2}\left(\Sigma^0-
\Sigma^z\right)=\left(\begin{array}{c|c}
    0 & 0 \\
    \hline
    0 & 1    \end{array}\right), \end{equation} where blocks are $1\times 1$
and $3\times 3$ in the case $\nu=1,3$ and $2\times 2$ in the case $\nu=2$. It
follows immediately that $H_G\ge 0$.

The matrices $N$ and $1-N$ can be viewed as projector operators that allow
only those rotations into the Hamiltonian (\ref{GoldHamGeneral}) that do
change the ground state. It is useful to understand what particular sub-set of
$SU(4)$ Lee group these physical rotations form. The vector field $\Omega_\mu$
can be expanded in the basis of fifteen generators of $SU(4)$ Lee group:
$\{\Sigma^l\}$, with $l=1..15$. And we subdivide them into two complementary
sets: the first one includes those generators that do commute with occupation
number matrices (\ref{Nfilled}, \ref{Nempty}), and we shall called it an
$(even)$ set, whereas the second one includes the remaining generators, and we
shall call it the $(odd)$ set. Generators of the $(even)$ set constitute an
algebra itself. This algebra has a normal Abelian subalgebra formed by a
single generator: $\Sigma^z-\mbox{tr}\Sigma^z/4$. A Lee group built around the
$(even)$ set of generators is called a stabilizer sub-group $S$ of $SU(4)$ Lee
group. The $(odd)$ set always contains an even number of generators.
Specifically, eight in the case of $\nu=2$ and six in the case $\nu=1,3$.

The Hamiltonian (\ref{GoldHamGeneral}) must be invariant under the time
reversal symmetry. The time reversal operator can be chosen as a complex
conjugate operator: $U\rightarrow U^*$. It follows that $\vec{\Omega}
\rightarrow -\vec{\Omega}^T$. Now it is evident that $\Omega_-$ and $\Omega_+$
get interchanged under the time reversal in (\ref{GoldHamGeneral}), and this
changes the Hamiltonian. But we remember that the time reversal is always
accompanied by inverting the magnetic field $B^z$ and, thus, we can restore
the time reversal symmetry by multiplying the antisymmetric part of the
Hamiltonian $H_G$ by the sign of the magnetic field:
\begin{equation}\label{GoldHam1} H_G={E_1\over2}\int{d^2\vec{r}\over 2\pi}
\left[\mbox{tr}\left((\Sigma^0-N)\Omega_\mu N\Omega_\mu\right)+i\,\mbox{sgn}
(B^z)\epsilon_{\mu\nu}\mbox{tr}\left(\Omega_\mu\Omega_\nu N\right)\right]
\end{equation}
Now if we define a useful gradient vector field:
\begin{equation}\label{OmegaZ} \Omega^z_\mu(\vec{r})={i\over 2}\mbox{tr}\left(
U^+(\vec{r})\Sigma^z\partial_\mu U(\vec{r})\right), \end{equation} then it is
straightforward to rewrite the Eq.(\ref{GoldHam1}):
\begin{equation}\label{GoldHam} H_G={E_1\over2}\int{d^2\vec{r}\over 2\pi}
\left[\mbox{tr}\left((\Sigma^0-N)\Omega_\mu N\Omega_\mu\right)+\mbox{sgn}(B^z)
\ \mbox{curl}\,\Omega^z\right], \end{equation} The first term
in $H_G$ (\ref{GoldHam}) is the gradient energy whereas the second term is
proportional to the topological index of an excited state:
\begin{equation}\label{Index}{\cal Q}=\int\mbox{curl}\,\Omega^z\ {d^2\vec{r}
\over 2\pi}= Z, \end{equation} where $Z$ is the set of integer numbers. The
case ${\cal Q}=\pm 1$ corresponds to the simplest spin skyrmion in the first
layer being rotated by a $SU(4)$ matrix to become a general bilayer skyrmion.
The energy constant in $H_G$ (\ref{GoldHam}) coincide identically with that of
the one-layer case \cite{IPF99}, which means that the bilayer skyrmion energy
is the same as found for one layer. But there is an important difference
between a bilayer skyrmion and a spin skyrmion in a single ferromagnetic
layer. Namely, as was shown in the Ref.\cite{IPF99} any skyrmion carries a
charge density in the core:
\begin{equation}\label{ChargeDens1}n(\vec{r})={\mbox{curl}\,\Omega^z(\vec{r})
\over2\pi}={R^2\over\pi\left(R^2+\vec{r}^2\right)^2}, \end{equation}
where $R$ is the radius of the skyrmion's core, contrary to that in the
bilayer a charge of the skyrmion's core can be delocalized over the two layers
with a much longer tails:
\begin{equation}\label{ChargeDens2}n_{1,2}(\vec{r})\sim\pm{1\over\left(
R^2+\vec{r}^2\right)}. \end{equation} The total charge of two layers does
converge according to (\ref{ChargeDens1}) at large distances from the
skyrmion's core.

The Goldstone Hamiltonian (\ref{GoldHam}) can be cast in a special order
parameter representation. To do this we define a non-homogeneous order
parameter matrix $Q$ as follows:
\begin{equation}\label{OrderQ}
Q(\vec{r})=U(\vec{r})NU^+(\vec{r}). \end{equation} This electronic order
parameter has an important property:
\begin{equation}\label{AverOrder} \langle A\rangle=\mbox{tr}(AQ),
\end{equation} where $A$ is any operator. Inspecting the particular
difinition of $N$ (\ref{Nfilled}) it becomes evident that rotations from
the denominator sub-group $S$ leaves the order parameter intact. Thus,
rotations in (\ref{OrderQ}) can be restricted to a coset or, in other word,
a physical space of the bilayer 2DEG:
\begin{equation}\label{Coset} {U(4)\over U(\nu)\otimes U(4-\nu)}.
\end{equation} Now it a straightforward calculation to rewrite $H_G$
(\ref{GoldHam}) in terms of the order parameter matrix:
\begin{equation}\label{SigmaModel} H={E_1\over 4}\int\,\mbox{tr}\left(
\vec{\nabla}Q\vec{\nabla}Q\right){d^2\vec{r}\over 2\pi}+\mbox{sgn}(B^z)
{E_1\over 2}\int\,\epsilon_{\mu\nu}\mbox{tr}\left(Q\partial_\mu Q\partial_\nu
Q\right){d^2\vec{r}\over2\pi}. \end{equation} In this representation the
topological index appears as an index of a map of the order parameter coset
space into a 2D plane. The index selection rule (\ref{Index}) is a consequence
of a well known homotopy group identity:
\begin{equation}\label{Pi2Coset}{\cal Q}=\pi_2\left({U(4)\over U(\nu)\otimes
U(4-\nu)}\right)=Z. \end{equation} In the end we have to include the Coulomb
energy of charge distribution inside the skyrmion core:
\begin{equation}\label{DirectCoulomb} \delta H_G={1\over 2}\int\int d^2
\vec{r}d^2\vec{r'}\ {\mbox{curl}\,\Omega^z(\vec{r})\over 2\pi}{e^2\over
|\vec{r}-\vec{r'}|}{\mbox{curl}\,\Omega^z(\vec{r}')\over 2\pi}.
\end{equation}

\section*{Anisotropic Part of Coulomb Energy. Phase Diagram.}

In this section we cast the anisotropic part of the bilayer Hamiltonian
(\ref{anisHam}) in terms of the order parameter matrix $Q$. It can be
conveniently done by the following Harteree-Fock average of $c$-operator
product in (\ref{anisHam}):
\begin{equation} \label{ProdAver} \begin{array}{c}\displaystyle
\tau^\mu_{\tau_1\tau_4}\tau^\nu_{\tau_2\tau_3}<c^+_{\alpha\tau_1p_1}
c^+_{\beta\tau_2p_2}c_{\beta\tau_3p_3}c_{\alpha\tau_4p_4}>=  \nonumber\\
\displaystyle =\delta_{p_1p_4}\delta_{p_2p_3}\mbox{tr}(Q\tau^\mu)\mbox{tr}
(Q\tau^\nu)-\delta_{p_1p_3}\delta_{p_2p_4}\mbox{tr}(Q\tau^\mu Q\tau^\nu),
\end{array} \end{equation} where $\tau^\mu$ acts on four-spinor as $\tau^\mu
\otimes\sigma^0$. Next, we define the following Coulomb anisotropy constants:
\begin{equation}\label{E_ab} E^{ab}=\int{dzd\bar{z}\over 2\pi l_H^2}V^{ab}
(|z|)\exp{-|z|^2\over 2l_H^2}\approx\int{dzd\bar{z}\over 2\pi l_H^2}
V^{ab}(|z|), \end{equation} where the last approximation holds for
$(ab)\ne(00)$ in the limit $d\ll l_H$. And, finally, we rewrite the anisotropy
Hamiltonian (\ref{anisHam}) in terms of order parameter matrix $Q$:
\begin{equation}\label{AnisHam}\begin{array}{c}\displaystyle  H^{anis}
/{\cal N}=-\left(t+(\nu-1)E^{0x}\right)\mbox{tr}\left(Q\tau^x\right)-\left(
\mu^z+(\nu-1)E^{0z}\right)\mbox{tr}\left(Q\tau^z\right)-\\ \displaystyle
-|g|\mu_BH\mbox{tr}\left(Q\sigma^z\right)+{1\over 2}E^{zz}\left[
\mbox{tr}\left(Q\tau^z\right)\mbox{tr}\left(Q\tau^z\right)-\mbox{tr}\left(Q
\tau^zQ\tau^z\right)\right], \end{array} \end{equation} where ${\cal N}$ is
the number of degeneracy of the Landau Level. The Eqs.(\ref{SigmaModel},
\ref{AnisHam}) defines the effective long-range Hamiltonian of a bilayer at
integer filling factors. At non-zero temperatures thermal fluctuations of the
order parameter soften the anisotropy constants in the Hamiltonian
(\ref{AnisHam}). The relevant calculation can be found in eg. \cite{ZJ93} and
the result reads:
\begin{equation}\label{Trenorm} \begin{array}{c}\displaystyle \left(t+(\nu-1)
E^{0x}\right)_R=\left(t+(\nu-1)E^{0x}\right)\left({l_h\over R^*}\right)^{8T
/E_1}\\ \displaystyle \left(t+(\nu-1)E^{0z}\right)_R=\left(t+(\nu-1)
E^{0z}\right)\left({l_h\over R^*}\right)^{8T/E_1}\\ \displaystyle
(|g|\mu_BH)_R=|g|\mu_BH\left({l_h\over R^*}\right)^{8T/E_1}\\ \displaystyle
E^{zz}_R=E^{zz}\left({l_h\over R^*}\right)^{24T/E_1},
\end{array} \end{equation} where the spatial scale $R_*^2=l_H^2E_1/\mbox{max}
(t,\mu^z,|g|\mu_BH,E^{zz})$ indicates the excitation wavelength where its
anisotropy energy starts to compete with its exchange energy. Note that the
three first constants renormalize as an external field whereas $E^{zz}$
constant renormalizes as an easy-axis anisotropy. Although the Coulomb energy
$E_1\sim 100K\gg T\sim 1K$ in most experiments, the specific number:
$24=3\times 8$, which is related to the order of anisotropy and to the eight
degree of freedom for thermal fluctuations in the case of $SU(4)$ symmetry,
makes the renormalization of the constant $E^{zz}$ noticeable.

As we have seen in the previous section the order parameter can be
parameterized by six or eight angles in the case $\nu=1,3$ or $\nu=2$.
Actually, not every of those rotations corresponds to a physically distinct
eigen state. We restrict the calculation of the total bilayer energy up to a
first order in powers of the anisotropy Hamiltonian, which means that we shall
need only its diagonal matrix elements. But, these are real matrix elements of
course, despite the fact that in an external magnetic field there is no time
reversal symmetry. Hence, if the Hamiltonian is real one so real has to be its
ground state. One generates all real eigen states from a reference state by
rotations from the $SO(4)$ sub-group of the $SU(4)$ group. This group has 6
parameters with two of them falling into the denominator sub-group. Thus, the
ground state differs from the reference state by just four rotations. One can
view locally the 8D manifold of order parameter as a composition of four unit
vectors: magnetization of the first and the second layers and the two
hopping-tau vectors which represent the distribution of spin-up and spin-down
electron density over the two layers. Now the first two term in the
Hamiltonian (\ref{anisHam}) are external fields acting on these four vectors.
On the other hand the Coulomb energy couples pairs of tau vectors via an
exchange interaction. This instructive picture allows us to identify only
three special global rotations that do change the total bilayer energy.
We start with the case $\nu=2$ and
we use a set of trial many electron wave functions parameterized by
the three angles of relevant in our case rotations: $\theta_\pm$ and
$\vartheta$,:
\begin{equation}\label{Trial}
\prod_p\ U(\vartheta,-\vartheta)R(\theta_+,\theta_-)\ c^+_{+1p}c^+_{-2p}\
|empty\rangle,         \end{equation}
where spins in the layer 1,2 are first rotated by $\pm\vartheta$:
\begin{equation}\label{U}
U^{\alpha\tau_1}_{\beta\tau_2}(\vartheta,-\vartheta)=\left({\tau^0+\tau^z\over
2}\right)_{\tau_1\tau_2}\exp(i{\vartheta\over 2}\sigma^y)_{\alpha\beta}+
\left({\tau^0-\tau^z\over 2}\right)_{\tau_1\tau_2}\exp(-i{\vartheta\over 2}
\sigma^y)_{\alpha\beta},       \end{equation}
and then wave functions of electrons with spin $\pm$ spill over the two
layers, the process described by two distinct angles: $\theta_\pm$,:
\begin{equation}\label{R}
R^{\alpha\tau_1}_{\beta\tau_2}(\theta_+,\theta_-)=\left({\sigma^0+\sigma^z\over
2}\right)_{\alpha\beta}\exp(i{\theta_+\over 2}\tau^y)_{\tau_1\tau_2}+\left({
\sigma^0-\sigma^z\over 2}\right)_{\alpha\beta}\exp(-i{\theta_-\over 2}
\tau^y)_{\tau_1\tau_2}.        \end{equation}
This set includes the singlet-liquid state at $\theta_\pm=\pi/2$ and
$\vartheta=0$ and the canted antiferromagnetic state at $\theta_\pm=0$. The
order parameter reads:
\begin{equation}\label{Ndens}Q=URNR^+U^+, \end{equation}
with $N$ being the electron density calculated with the reference state of the
previous section (see (\ref{Nfilled})). Now we substitute (\ref{Ndens}) into
the anisotropic Hamiltonian (\ref{AnisHam}) to find the the total anisotropy
bilayer energy as:
\begin{equation}\label{Total}        \begin{array}{c}\displaystyle
E^{anis}=-E^{zz}\cos\theta_+\cos\theta_--(t+E^{0x})\cos\vartheta(\sin\theta_+
+\sin\theta_-)- \\ \displaystyle -(\mu^z+E^{0z})(\cos\theta_+-\cos\theta_-)-
|g|\mu_BH\sin\vartheta(\cos\theta_++\cos\theta_-), \end{array} \end{equation}
The minimum of this energy corresponds to three phases: a) ferromagnetic
$\vartheta=\pi/2$, $\theta_+=\theta_-=0$; b) spin singlet $\vartheta=0$,
$\theta_+=\pi-\theta_-=\theta$; and c) canted antiferromagnetic state
otherwise, as it is shown on the Fig.3. It is identical to that found in the
Ref.\cite{DD99}. A line of continuous phase transitions between the
ferromagnetic phase and the canted antiferromagnetic phase is given by the
following equation:
\begin{equation}\label{FAPhT}
\left[\left(E^{zz}+|g|\mu_BH\right)^2-\left(\mu^z+E^{0z}\right)^2\right]
|g|\mu_BH=\left(t+E^{0x}\right)^2\left(E^{zz}+|g|\mu_BH\right)
\end{equation}
In the spin singlet phase the interlayer mixing phase: $\theta$, is determined
by the equation:
\begin{equation}\label{theta}
\left(E^{zz}\sin\theta+t+E^{0x}\right)\cos\theta=\left(\mu^z+E^{0z}\right)
\sin\theta .   \end{equation}
A phase transition line that separate the spin singlet phase from the canted
antiferromagnetic phase is given parametrically by the equation:
\begin{equation}\label{SAPhT}
\left((t+E^{0x})\sin\theta-E^{zz}+\left(\mu^z+E^{0z}\right)\cos\theta\right)
\left(t+E^{0x}\right)=\left(|g|\mu_BH\right)^2\sin\theta ,
\end{equation}
where $\theta$ is determined from (\ref{theta}). This phase transition is a
continuous one also.

\begin{figure}\label{PhaseDiag}
\epsfxsize=4.0in
\centerline{\epsffile{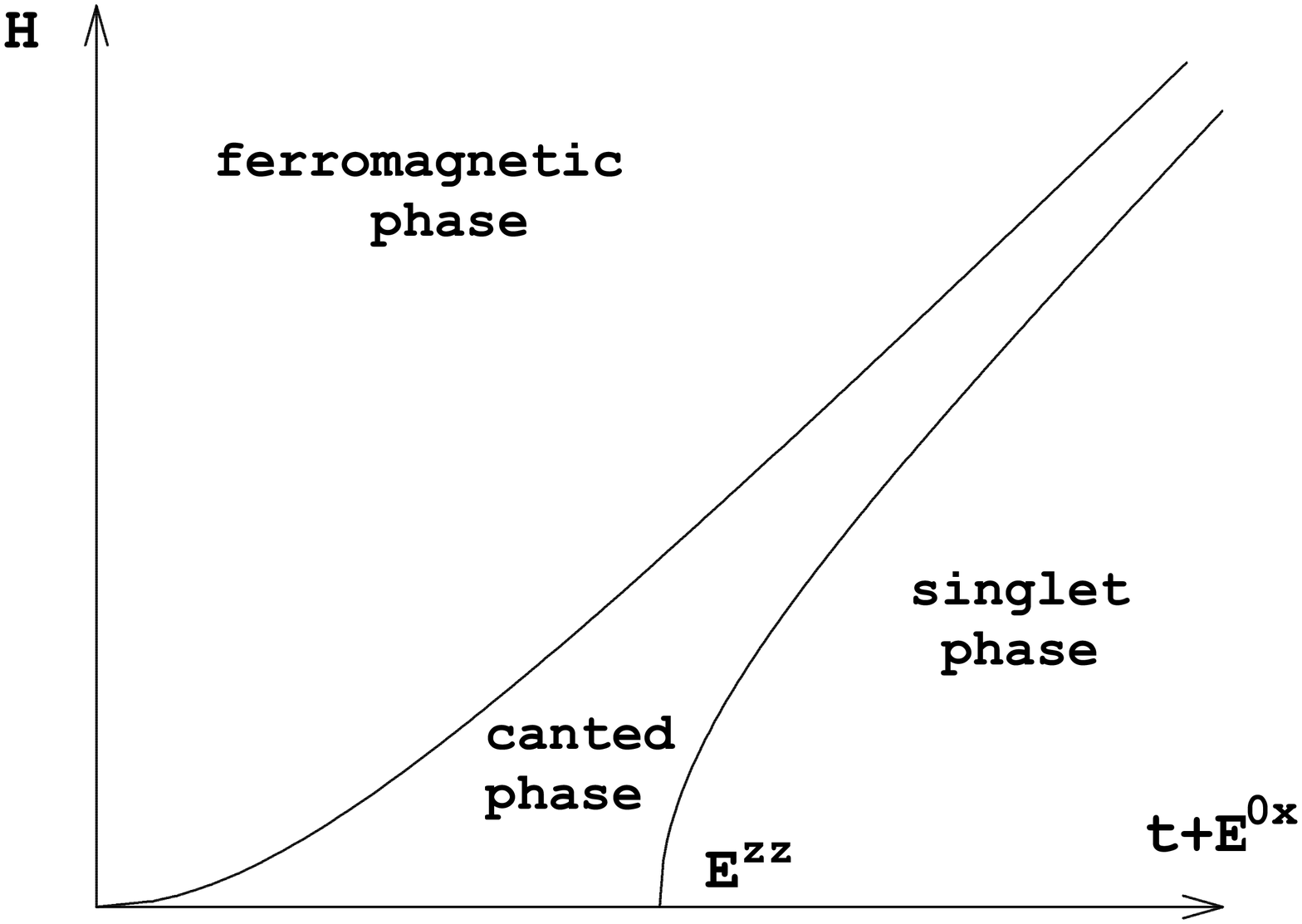}}
\caption{Phase Diagram in the Gate-symmetric case} \end{figure}

In the case $\nu=1$ there is no Coulomb interaction energy and the total
bilayer energy reads:
\begin{equation}\label{Nu1Total} E^{anis}
=-t\sin\theta-\mu^z\cos\theta-|g|\mu_BH\cos\vartheta, \end{equation}
The minimum of this energy is given by electron spin being directed along the
magnetic field: $\vartheta=0$, whereas $\theta=\tan^{-1}t/\mu^z$. There is no
phase transition in the case $\nu=1$ and the only phase can be characterized
as ferromagnetic in both the spin and the layer spaces.

The case $\nu=3$ formally reduces to the case $\nu=1$ although here the
Coulomb interaction energy does not vanish identically. We find
renormalizations to the one-particle electron Hamiltonian whereas the total
energy being similar to the case $\nu=1$:
\begin{equation}\label{Nu3Total}
E^{anis}=-(t+2E^{0x})\sin\theta-(\mu^z+2E^{0z})\cos\theta- 
|g|\mu_BH\cos\vartheta, \end{equation}
There is no phase transition in this case either.

\section*{Anisotropic energy gap of one Skyrmion}

In this section we find an anisotropic part of the total skyrmion gap energy.
The non-homogeneous order parameter that represents one skyrmion is given by
the Belavin-Polyakov (BP) skyrmion solution for $|{\cal Q}|=1$ \cite{BP75}:
\begin{equation} \label{SkyRotat}
Q_{BP}(z\bar{z})= {R^2\over {R^2+|z|^2}}\left( \begin{array}{c|c}
    |z|^2 & zR \\
    \hline
    \bar{z}R & R^2
  \end{array}  \right)       \end{equation}
where only the shown above four matrix elements differs from those in the
electron density matrix $N$ (\ref{Nfilled}). The skyrmion order parameter has
to be rotated by a homogeneous matrix $RU$ calculated in the previous section
in such a way that the order parameter far away from the skyrmion center
minimizes the anisotropy energy. In addition to this rotation we have to allow
all homogeneous rotations from the denominator sub-group $S$ that actually
transform the BP skyrmion order parameter (\ref{SkyRotat}): $W$. Thus, a
general skyrmion order parameter reads:
\begin{equation}\label{BelRot}
Q(\vec{r})=RUWQ_{BP}(z\bar{z})W^+U^+R^+. \end{equation} First, we consider
the case $\nu=2$. Here the matrix $W$ is parameterized by seven angles:
\begin{equation}\label{Wdenom} \left( \begin{array}{cccc} \displaystyle \cos
\beta_f e^{i(\gamma_f+\alpha_f)} & \sin\beta_f e^{i(\gamma_f-\alpha_f)} & 0 &
0 \\ \displaystyle -\sin\beta_f e^{i(-\gamma_f+\alpha_f)} & \cos\beta_f
e^{i(-\gamma_f-\alpha_f)} & 0 & 0 \\ \displaystyle 0 & 0 & \cos\beta_e
e^{i(\gamma_e+\alpha_e)} & \sin\beta_e e^{i(\gamma_e-\alpha_e)} \\
\displaystyle 0 & 0 & -\sin\beta_e e^{i(-\gamma_e+\alpha_e)} & \cos\beta_e
e^{i(-\gamma_e-\alpha_e)} \end{array} \right) \end{equation} The additional
seventh parameter angle of the denominator group just rotates the coordinates:
$z\rightarrow e^{i\gamma_7}z$. We find by explicit calculation that the
skyrmion anisotropic energy does not depend on parameters $\gamma_e$,
$\gamma_f$ and $\gamma_7$ whereas $\alpha_e=0$ and $\alpha_f=\pi$ correspond
to the skyrmion energy minimum. Thus we shall express the anisotropic skyrmion
energy in terms of the two relevant parameters $\beta_e$ and $\beta_f$ that
rotates the core of skyrmion in the empty-electron space and filled-electron
space correspondingly. Besides these angles the skyrmion energy depends on
free parameters entering the BP solution. The calculation here has been
performed only for skyrmion topological index: $|{\cal Q}|=1$ (\ref{SkyRotat})
and in this case there is only one such parameter, namely, the radius $R$ of
the core of skyrmion. There are few different spatial integrals that we
encounter in calculation. And only one of them is logarithmically divergent.
We calculate the skyrmion anisotropy energy with the logarithmic accuracy. It
means that parts of the anisotropy energy coming from different anisotropy
sources are all multiplied by the same spatial integral, which we denote by a
constant $A$:
\begin{equation}\label{Acons}  A=
\left({R\over l_H}\right)^2\log{R^*\over R}. \end{equation} where $R^*$ is the
inverse mass of Goldstone excitations in the model (\ref{GoldHam},\ref{Total}).
Also, the Zeeman energy of skyrmion reads:
\begin{equation}\label{Nu2Zeeman} \begin{array}{c}\displaystyle
E^{skyr}_Z=A{\cal E}_Z=A|g|\mu_BH\left(-\ {1\over2}\sin\vartheta(\cos\beta_f-
\cos\beta_e) (\cos\theta_+-\cos\theta_-)+\right. \\ \displaystyle \left.
+\sin\vartheta(\cos\theta_++\cos\theta_-)-\cos\vartheta(\sin\beta_f+\sin
\beta_e)\sin{\theta_++\theta_- \over2} \right)\end{array} \end{equation} The
hopping energy of skyrmion reads:
\begin{equation}\label{Nu2Hopping} \begin{array}{c}\displaystyle
E^{skyr}_H=A{\cal E}_H=A(t+E^{0x})\left(-\ {1\over2}\cos\vartheta(\cos\beta_f
-\cos\beta_e)(\sin\theta_+-\sin\theta_-)+\right. \\ \displaystyle\left. +\cos
\vartheta(\sin\theta_++\sin\theta_-)-\sin\vartheta(\sin\beta_f+\sin\beta_e)
\cos{\theta_++\theta_- \over2} \right)\end{array} \end{equation} The gate
asymmetry energy of skyrmion reads:
\begin{equation}\label{Nu2GateAsym} \begin{array}{c}\displaystyle
E^{skyr}_G=A{\cal E}_G=A(\mu_z+E^{0z})\left((\cos\theta_+-\cos\theta_-)-
\right. \\ \displaystyle \left. \ {1\over2}(\cos\beta_f-\cos\beta_e)
(\cos\theta_++\cos\theta_-) \right)\end{array} \end{equation} And finally the
anisotropic Coulomb energy of skyrmion reads:
\begin{equation}\label{Nu2CouSky} \begin{array}{c}\displaystyle
E^{skyr}_{zz}=A{\cal E}_{zz}=AE^{zz}\left(-\ {1\over2}(1+\cos\beta_f\cos\beta_
e)(1+\cos\theta_+\cos\theta_-)+\right. \\ \displaystyle \left. +2\cos
\theta_+\cos\theta_--\ {1\over2}\sin\beta_f\sin\beta_e\sin\theta_+\sin\theta_-
\right).\end{array} \end{equation} There is also a contribution to the
skyrmion energy coming from an non-homogeneous BP electric charge distribution
inside the skyrmion core (\ref{DirectCoulomb}):
\begin{equation}\label{CouCharge}
E^{skyr}_C={1\over 2}\int\int d^2\vec{r}d^2\vec{r'}{R^2\over\pi (r^2+R^2)^2}
{e^2\over \kappa |\vec{r}-\vec{r'}|}{R^2\over\pi (r'^2+R^2)^2}={3\pi^2\over
64} {e^2\over \kappa R}. \end{equation} The minimum of the total anisotropic
skyrmion energy (\ref{Nu2Zeeman},\ref{Nu2Hopping},\ref{Nu2GateAsym},
\ref{Nu2CouSky}): ${\cal E}^{sky}={\cal E}_Z+{\cal E}_H+{\cal E}_G+{\cal
E}_{zz}$, over the two parameters $\beta_e$ and $\beta_f$ was found
numerically and is denoted as: ${\cal E}^{skyr}_{min}$. Next, we find a
minimum the total skyrmion energy including (\ref{CouCharge}):
$E^{skyr}=A{\cal E}^{skyr}_{min}+E^{skyr}_C$, with respect to the skyrmion
radius $R$:
\begin{equation}\label{GapSk}
\Delta={{\cal Q}+|{\cal Q}|\over 2}E_1+3\left({\cal E}^{skyr}_{min}\left(
{3\pi^2e^2\over 128\kappa l_H}\right)^2\log{e^2\over \kappa l_H{\cal
E}_{min}^{skyr}}\right)^{1/3}. \end{equation} This formula holds in the limit
${\cal E}^{skyr}\ll E_1$ and, thus, the second term is much smaller then the
first term in the Eq.(\ref{GapSk}) as they are calculated. But, it is
important in the case of antiskyrmion ${\cal Q}=-|{\cal Q}|$, the gap is only
a relatively small anisotropic energy. The resulting anisotropic part of a
skyrmion gap is shown on the Fig.4 in the case $\mu^z+E^{0z}=0$. Note the two
prominent cusp-like lines in the skyrmion gap sheet coincide with the two
phase transition lines from the Fig.3. A view of another cross-section of the
skyrmion gap: $|g|\mu_BH=0$, is shown on the Fig.5. A minima here also 
coincide
with the canted-antiferromagnetic phase. A skyrmion in the ferromagnetic state
is a spin-skyrmion with the spin rotations being localized inside one of the
two layer, whereas a skyrmion in the spin-singlet state is a layer-skyrmion
with the electron density being distributed over the two layers.

In the experimental setup \cite{K00} ${\cal E}^{skyr}\approx E_1$ and our
formulas can be compared with the experimental results only qualitatively. But
they found a profound disappearence of the thermal activation gap maximum at
some interval on the $\nu=2$ line. In our theory this would correspond to a
minimum in the skyrmion gap and we suggest that this indicate the canted
antiferromagnetic phase.

\begin{figure}\label{SkyGap}
\epsfxsize=6.0in
\centerline{\epsffile{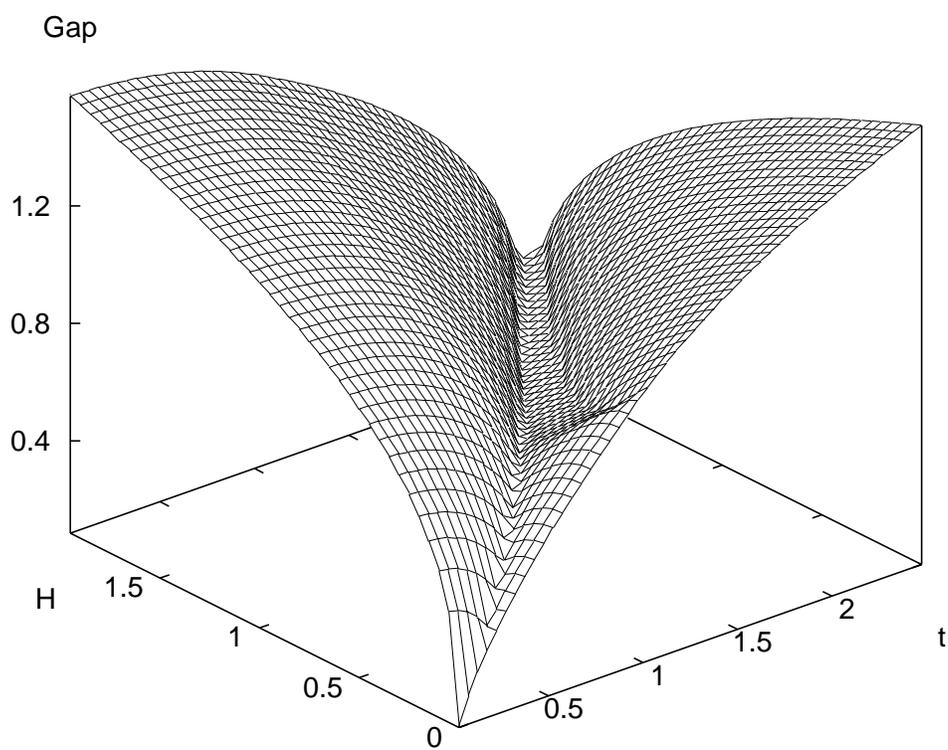}}
\caption{Anisotropic part of Skyrmion Energy for a vanishing gate asymmetry.}
\end{figure}
\begin{figure}\label{SkyGap1}
\epsfxsize=6.0in
\centerline{\epsffile{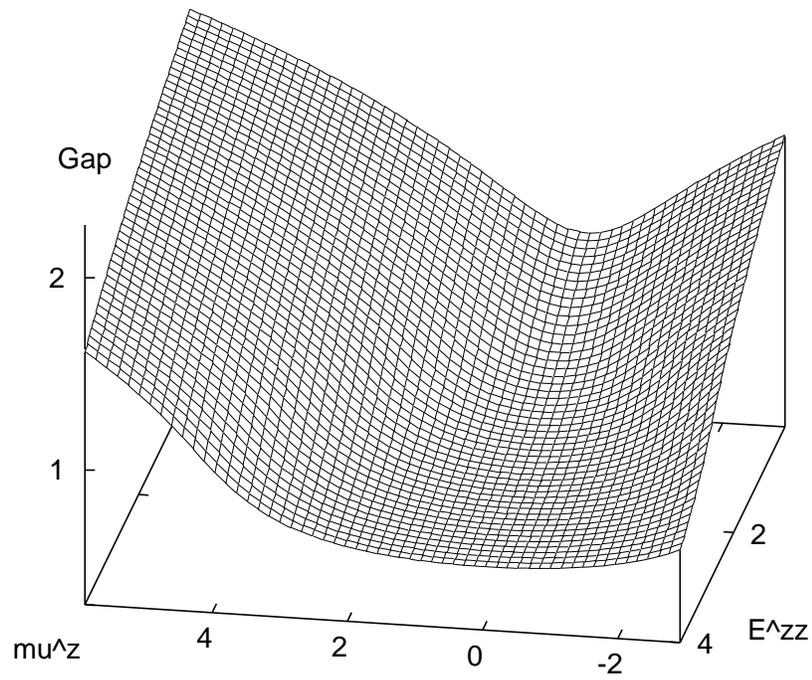}}
\caption{Anisotropic part of Skyrmion Energy in the case of vanishing
Zeeman energy. Spin-singlet phase is on the right side and the canted
antiferromagnetic phase is on the left side} \end{figure}

In the case $\nu=1$ we parameterize general rotations from the denominator
sub-group by four angles in such a way that the electron density matrix
becomes:
\begin{equation} \label{Nu1Dens}
WQ_{BP}(z\bar{z})W^+= {1\over{R^2+|z|^2}}\left( \begin{array}{c|c}
    |z|^2 & Rz|f\rangle \\
    \hline
    R\bar{z}\langle f| & R^2|f\rangle\langle f|
  \end{array}  \right)       \end{equation} where
\begin{equation}\label{Fket}
|f\rangle=\left (\cos{\beta\over 2},\sin{\beta\over 2}\cos\alpha
e^{i\lambda_1},\sin{\beta\over 2}\sin\alpha e^{i\lambda_2}\right)
\end{equation} Once again we find that the skyrmion energy does not depend
on the parameters $\lambda_1$ and $\lambda_2$. A straightforward calculation
shows then that the Zeeman skyrmion energy is
\begin{equation}\label{Nu1Zeeman}
\Delta=2A|g|\mu_BH\left(1-\sin^2{\beta\over 2}\cos^2\alpha\right),
\end{equation}
the Hopping skyrmion energy is
\begin{equation}\label{Nu1Hopping}
E^{skyr}_H=At\left(-\sin\beta\sin\alpha+\sin\theta\left[1+\cos^2\alpha\sin^2
{\beta\over 2}\right]\right) \end{equation} whereas the gate asymmetry
skyrmion energy is:
\begin{equation}\label{Nu1GateAsym}
E^{skyr}_G=A\mu_z\left(-\cos\beta\left[1-\sin^2{\beta\over 2}\cos^2
\alpha\right]-\cos^2\alpha\sin^2{\theta\over 2}+\cos\theta\right)
\end{equation}
Note that there is no Coulomb energy in the case $\nu=1$. Searching for minima
of Eq.(\ref{Nu1Zeeman},\ref{Nu1Hopping},\ref{Nu1GateAsym}) varying parameters
$\beta$ and $\alpha$ and fixing $\theta=\tan^{-1}t/\mu^z$ we find the minimal
anisotropy energy of skyrmion to be:
\begin{equation}\label{Nu1Anis}
\Delta={{\cal Q}+|{\cal Q}|\over 2}E_1+\mbox{min}\left(2\sqrt{t^2+\mu_z^2}\
,\ 2|g|\mu_BH\right). \end{equation} The skyrmion gap is given by the same
formula as in the case $\nu=2$: (\ref{GapSk}). Generally, there is no
prominent minima in the skyrmion gap here. Nevetherless in the limit $t\ll
|g|\mu_BH$ such a minimum occurs.

The case $\nu=3$ reduces to the case $\nu=1$ if one simply to renormalize the
constant in (\ref{Nu1Anis}) as it was explained in the previous section.

\section*{Conductivity Activation Energy}

The conductivity activation energy was measured in experiments for bilayer
system \cite{K00} and in one layer system \cite{??}. It was known for quite
some time that this energy is considerably less than a typical exchange
constant $e^2/\kappa l_H$. This fact was the main motivation for the
experimental search of topological excitations. If one considers an act of
creation of skyrmion and anti-skyrmion pair at large separation then one
easily gets the pair energy: $E_1=1/2E_0$. This is definitely less than the
creation of electron-hole pair at large separation: $E_0$, but still of the
order of $e^2/\kappa l_H$, for one layer. In a bilayer system the prominent
minimum in the minimal activation energy for skyrmion antiskyrmion pair is
still of the order of $E_1$, in spite of the fact that experimental
conductivity activation energy goes to zero very sharply \cite{K00}.

This controversy can be overcome by considering the creation of neutral
antyskyrmion and electron pair at large separation. First, we consider a one
layer case. The energy of the additional electron with reversed spin does not
contain exchange Coulomb energy and consist from anisotropy energy only. The
energy of anti-skyrmion also has only anisotropic energy (\ref{GapSk}). The
anisotropic energy of electron can be neglected in the limit of large ratio:
$E_1/|g|\mu_BH$, which holds in most experiments. The total anti-skyrmion 
energy is:
\begin{equation}\label{AntiSkEn}
|g|\mu_BB\int(1-\cos\beta(r)){d^2\vec{r}\over 2\pi l_H^2}+{1\over 2}\int{e^2
\over\kappa|\vec{r}-\vec{r}'|}\mbox{curl}\,\Omega^z(\vec{r})\mbox{curl}\,
\Omega^z(\vec{r}')d^2\vec{r}\, d^2\vec{r}'. \end{equation}. Plugging the
BP solution we calculate this energy to be:
\begin{equation}\label{AnSkEn} 2|g|\mu_BBR^2\log{R^*\over R}\ +{3\pi^2\over 
64}{e^2\over\kappa R}. \end{equation} Minimizing it further with respect to
radius of the anti-skyrmion core $R$ we find the core radius:
\begin{equation}\label{RadSk}R={1\over 2}l_H\left({3\pi^2e^2\over 32\kappa l_H
|g|\mu_B\log(R^*/l_H)}\right)^{1/3} \end{equation} and the activation energy:
\begin{equation}\label{GapSkSing} \Delta={3\over 2}\left(|g|\mu_BH\log(R^*/l_H)
\left({3\pi^2e^2\over 128\kappa l_H}\right)^2\right)^{1/3}. \end{equation}
The upper limit under the logarithm $R^*$ is defined by the validity of BP
solution. Essentially, it is a distance where gradient energy becomes of the
order of Zeeman energy:
\begin{equation}\label{R^*} R^*=l_H\sqrt{E_1\over |g|\mu_BH}. \end{equation}
Therefore the logarithmic factor is of the order of unity. If it were rather
large one would need to compare the energy of ${\cal Q}=-1$ and ${\cal Q}=-2$
antiskyrmions, with the latter being logarithmic free. In any case an
important point is that the activation energy is magnetic field dependent:
\begin{equation}\label{DeltaField} \Delta=K H^{2/3}, \end{equation} which in
agreement with the various experimental results \cite{??}. The absolute value
of the constant $K$ (\ref{GapSkSing})in this relation also conforms experiment
\cite{??}. Note that the Zeeman energy of electron is much less that $\Delta$
(\ref{DeltaField}). In the case of double layer system the situation is more
complicated due to large number of parameters defining the $H^{anis}$.
Nevertheless, assuming anisotropy energy is small and neglecting for the same
reason the anisotropy energy of an electron we get the deep minimum in the
canted antiferromagnetic phase in accordance with the experimental result
\cite{K00}. It should be noted that in the process of creation of
anti-skyrmion electron pair the total topological charge of the system is not
conserved as opposed to the skyrmion antiskyrmion pair. Therefore, this
process goes tentatively on the sample boundary. Also the existence of
magnetic field makes it possible to violate electron-hole symmetry usually
assumed in theories used the basis of projected on the lowest Landau Level
wave functions. The violation of the electron-hole symmetry is related to the
topological charge and gives rise to the skyrmion vs antiskyrmion energy
difference due to the topological term in the action (\ref{GoldHam}).

\section*{Acknowledgement}

Valuable discussions with V.T.Dolgopolov and A.Maltsev are gratefully
acknowledged. This work was supported by grant INTAS N97-31980.

\end{document}